%% file: vis.tex
\documentclass[journal]{vgtc}                     

\onlineid{1495}

\vgtccategory{Research}

\vgtcpapertype{evaluation}

\title{Evaluating judgment of spatial correlation in visual displays of scalar field distributions}

\author{%
  Yayan Zhao and
  Matthew Berger
}

\authorfooter{
  \item Yayan Zhao and Matthew Berger are with Vanderbilt University. E-mail: yayan.zhao@vanderbilt.edu
}

\abstract{%
In this work we study the identification of spatial correlation in distributions of 2D scalar fields, presented across different forms of visual displays. We study simple visual displays that directly show color-mapped scalar fields, namely those drawn from a distribution, and whether humans can identify strongly correlated spatial regions in these displays. In this setting, the recognition of correlation requires making judgments on a set of fields, rather than just one field. Thus, in our experimental design we compare two basic visualization designs: animation-based displays against juxtaposed views of scalar fields, along different choices of color scales. Moreover, we investigate the impacts of the distribution itself, controlling for the level of spatial correlation and discriminability in spatial scales. Our study's results illustrate the impacts of these distribution characteristics, while also highlighting how different visual displays impact the types of judgments made in assessing spatial correlation. Supplemental material is available at \url{https://osf.io/zn4qy/}
}


\teaser{
\centering
\includegraphics[width=0.9\linewidth]{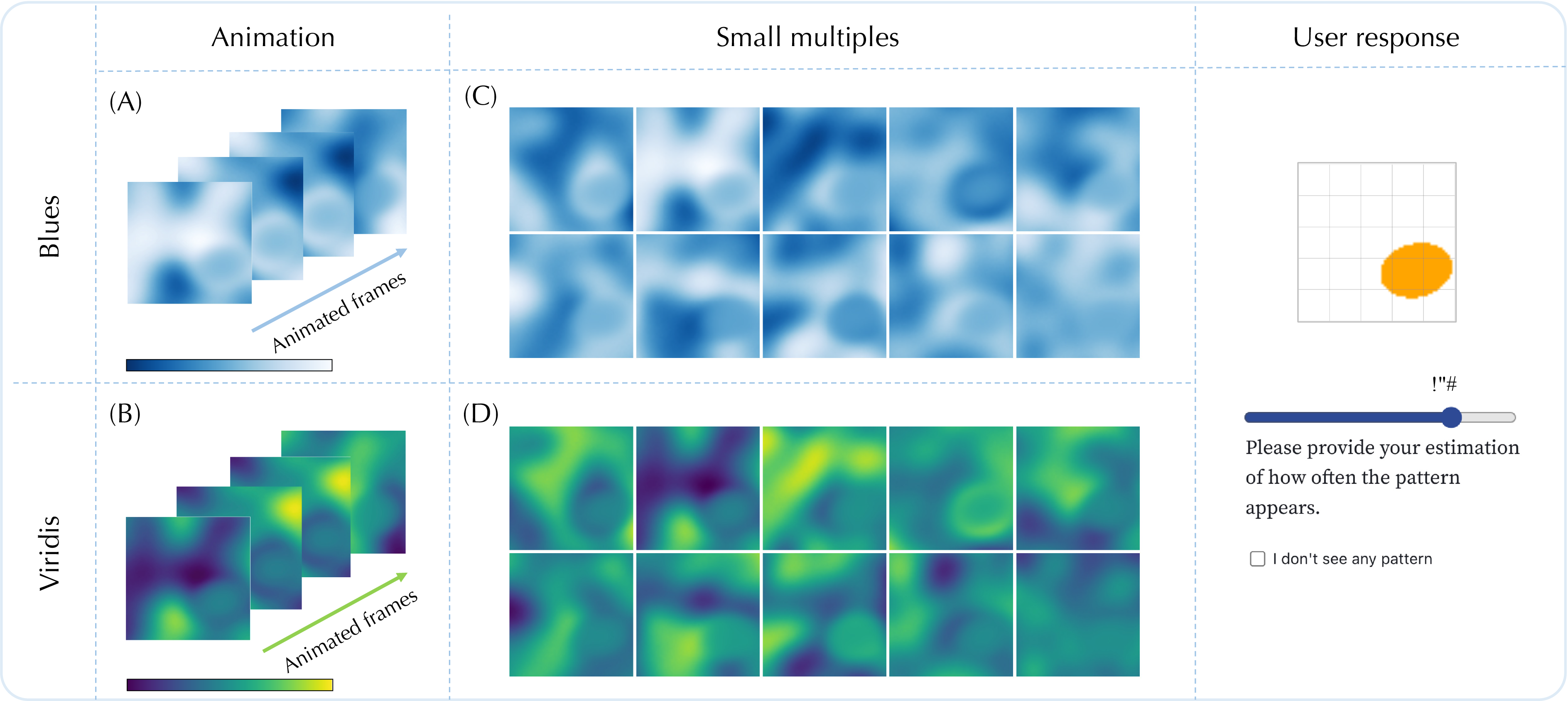}
    \caption{We study different visualization designs of 2D scalar field distributions for the task of spatial correlation perception. We compare static (small multiples) versus dynamic (animation) visualization types, as well as multi-hue versus single-hue luminance-varying color scales. We study the frequency with which participants observe views that give evidence of correlation, as well as their accuracy in identifying the target correlated region.
     }
\label{fig:teaser}
}

\graphicspath{{figs/}{figures/}{pictures/}{images/}{./}} 

\usepackage{tabu}                      
\usepackage{booktabs}                  
\usepackage{lipsum}                    
\usepackage{mwe}                       
\usepackage{amssymb}
\usepackage{amsmath}
\usepackage{siunitx}

\usepackage{mathptmx}                  

\begin{document}

\definecolor{midblue}{rgb}{0.3865, 0.6395, 0.798}
\definecolor{lightblue}{rgb}{0.651, 0.808, 0.890}


\firstsection{Introduction}
\label{sec:intro}

\maketitle

\input{sections/intro}

\section{Related Work}
\label{sec:related}
\input{sections/related}

\section{Background and Hypothesis}
\label{sec:background_hypothesis}
\input{sections/background}

\section{Experimental Design}
\label{sec:study_design}
\input{sections/study_design}

\section{Results and Findings}
\label{sec:result}
\input{sections/result}

\section{Discussion}
\label{sec:discussion}
\input{sections/discussion}

\bibliographystyle{abbrv-doi-hyperref}

\bibliography{template,new_refs}

\end{document}

%% file: sections/intro.tex
Visualization is at its best when perception of visualized data facilitates meaningful analysis of data.
Towards this end, empirical studies within the visualization community have provided guidelines on what types of visualization designs work well for particular analytical tasks~\cite{srabanti2022comparative, hullman2015hypothetical,hullman2018pursuit, kale2018hypothetical,zhang2021visualizing}.
Among these tasks, the matter of \emph{correlation} has been extensively studied~\cite{pfaffelmoser2012visualization, ferstl2016visual, beecham2016map}, and specifically, the question of whether or not correlation exists within a visual display of data.
Visualization effectiveness of correlation has primarily been investigated for pairs of variables, namely the task of determining whether two variables are related to one another, although more complex data, e.g. spatial autocorrelation in choropleth maps~\cite{beecham2016map}, has been considered.
These studies shed light on effectiveness of different types of visual encodings~\cite{beecham2016map, whitaker2013contour}, and the impacts of different spatial structures used in visual encodings~\cite{pfaffelmoser2012visualization, ferstl2016visual}.

Previous studies of correlation focus on relatively simple data sources, in turn leading to simple, often static and single-view, displays.
However, correlation is not limited to these settings.
In particular, for scalar field-based data, it is common to encounter a \emph{distribution} of fields~\cite{um2021spot, bykov2021explaining, sanyal2010noodles, hollt2014ovis}, and correlation generalizes from just a pair of variables, to a set of variables.
In particular, when highly correlated variables are grouped together in a contiguous spatial region, this gives rise to \emph{spatial correlation}~\cite{banerjee2008gaussian}.

Spatial correlation analysis is of interest in numerous domains that yield scalar field distributions. In one scenario, we might be provided fields drawn from a distribution, from which strongly correlated spatial regions help illustrate structural variability. For instance in climate science~\cite{reckinger2015study}, it is common to run numerical simulations controlled by initial conditions of some physical quantity, such as viscosity. In turn, the simulation output gives a scalar field depicting a forecast of some numeric quantity, e.g. temperature. Running an ensemble of simulations over the parameter space gives a collection of fields, from which a region of strong spatial correlation can highlight structures that are robust to changes in parameters of interest~\cite{pfaffelmoser2012visualization}. In another scenario, we might build a probabilistic model for prediction over a spatial domain, provided scattered data measurements as input, wherein correlated regions can help in analyzing epistemic (model) uncertainty. Gaussian processes~\cite{williams2006gaussian} are common in this setting, whereby posterior inference leads to a distribution of scalar fields~\cite{wilson2020efficiently}. Identifying the existence of highly correlated regions is helpful in numerous areas, e.g. observing consistent/spurious structures within probabilistic surface reconstruction~\cite{sellan2022stochastic,holalkere2025stochastic}, or in the context of Bayesian optimization, analyzing the effects of covariance for adaptive data acquisition~\cite{maddox2021bayesian}.
In the visualization community, numerous techniques have been proposed to extract strongly correlated spatial regions~\cite{pfaffelmoser2012visualization, whitaker2013contour}, and more broadly, summaries of a distribution via generalized notions of spatial means and covariances~\cite{ferstl2015streamline, ferstl2016visual, pont2021wasserstein}, with the extracted spatial quantities subsequently displayed for users in support of visual analysis.


In this paper, we are interested in studying human perception of correlation in 2D scalar field distributions.
However, we depart from existing techniques~\cite{pfaffelmoser2012visualization, ferstl2016visual} that make assumptions on what constitutes strongly correlated regions.
Akin to prior perception studies~\cite{beecham2016map, whitaker2013contour}, we wish to see whether humans can perceive correlation using simpler, and more direct, visualizations of scalar fields, with a single field visualized as a color-mapped image.
Yet we must contend with a more complicated source of data: it is necessary to show a \emph{set} of scalar fields, rather than just one, in order to make a judgment of correlation.


We contribute a user study to test the effectiveness of visualization designs in helping users make decisions on correlation, within the problem setting of scalar fields sampled from a common distribution.
Please see Fig.~\ref{fig:teaser} for an overview of our approach.
We study visualizations along two main axes: the type of visual display, and choice of color scale.
Specifically, we compare animation-based displays, in which visualized scalar fields are shown as different frames in an animation, against small multiples displays, where fields are spatially arranged as a set of juxtaposed views.
Moreover, we study two types of luminance-increasing color scales: one in which the hue is fixed, and a multi-hue color ramp.
Our data is conceived via the construction of scalar field distributions in which a predefined spatial region is prescribed to have a specific level of correlation.
In contrast, the rest of the field is distributed under a Gaussian process prior~\cite{mackay1998introduction} of prescribed length scale, in order to give rise to spatial structures of varying size.
We study whether humans are able to perceive these correlated regions, while also eliciting their judgment on \emph{where} correlation exists.

We summarize the findings of our study:
\begin{itemize}
\item We find that small multiples leads to increased recognition of correlation over animation. This effect is most pronounced when the discriminability level is smallest.
\item In eliciting judgments on where strongly correlated regions exist, we find animation gives improved accuracy over small multiples. This is expected, since in an animation-based display, there is no need to spatially align views.
\item However, we find that this difference is primarily due to a \emph{translation} in the user's specified region, rather than a poor judgment on \emph{shape}, represented as orientation and size.
\item The choice of color scale impacts both recognition and specification of correlated regions. Multi-hue color ramps lead to increased recognition, yet also lead to users \emph{underestimating} the specific region where correlation exists.
\end{itemize}
Our findings suggest that simple visualization design choices can be effective for making inferences on sets of scalar fields.
Nevertheless, our study highlights conditions in which people face difficulties in making correlation judgments; we believe these conditions can point towards future work on dedicated methods for extracting, and visualizing, correlated regions.

%% file: sections/related.tex
Our work is most closely related to (1) visualization of correlation, (2) color perception in visualization design, (3) uncertainty visualization, and (4) comparison between animation and small multiples. We discuss each in turn.

\subsection{Visualization of correlation}
\label{sec:rw_corr}

The visualization of correlation has been extensively studied in recent years \cite{pfaffelmoser2012visualization, ferstl2016visual, beecham2016map}.
For example, Hullman et al.~\cite{hullman2015hypothetical} studied the effectiveness of a hypothetical outcome plot in visualizing the correlation between pairs of variables. 
Many studies focus on more complex data such as multi-dimensional scalar fields or geographic maps.
For example, Ferstl et al.~\cite{ferstl2016visual} calculated the correlation between occurrences of iso-contours of multidimensional scalar fields, and visualized the derived statistical properties through a variant of variability plots for streamlines.
Pfaffelmoser et al.~\cite{pfaffelmoser2012visualization} calculated global correlation structures on a distribution of 2D scalar fields, and embedded the correlation information into visualizations of other statistical quantities, namely the mean and standard deviation, via cluster coloring. 

Existing methods approach the visualization of correlation in one of two ways.
They derive statistical information such as correlation structure from the data, and visualize the derived information on top of the data.
Alternatively, they evaluate the effectiveness of a given visualization design, namely, whether humans can recognize the existence of correlation on a visual display of data. 
For example, many works \cite{harrison2014ranking, kay2015beyond} quantified how humans perceive correlation of pairs of variables in different visualization types.
Beecham et al.~\cite{beecham2016map} conducted a set of crowdsourced experiments to determine the just noticeable difference (JND) between pairs of choropleth maps of geographic units controlling for spatial autocorrelation and geometric configuration (variance in spatial unit area).

\subsection{Color perception in visualization design}
The focus of our work is closely related to color perception, in relation to visualization design.
A traditional guideline design is to favor expressive visualization choices~\cite{munzner2014visualization}, and for encoding quantitative data with color, this usually amounts to favoring luminance-varying color scales.
These scales adjust color brightness in correspondence with data magnitude, aligning with the natural human perception that interprets lighter colors as higher values and darker colors as lower values.
Nevertheless, recent work has challenged this convention, largely for color discriminability tasks~\cite{bujack2017good,szafir2017modeling}, as well as tasks pertaining to pattern discrimination~\cite{reda2020rainbows,reda2022rainbow}, identifying that color scales need not strictly vary by luminance.
Moreover, for color similarity tasks, further discernment along luminance-varying color scales indicate multi-hue scales are more effective than single-hue scales~\cite{liu2018somewhere}.
The benefits and costs of multi-hue compared to single-hue in color scales for different inference tasks has been studied in detail\cite{reda2022rainbow}.
It was found that for color scales which have high color categorization such as rainbow, diverging, or multi-hue, humans have a bias of detecting changes in global structure of scalar fields, but at the cost of reduced sensitivity to features defined by small, localized changes.
Yet, for color scales with low color categorization such as single-hue, humans tend to focused on localized feature defined by small variations.



We are interested in the effectiveness of different color scales in how humans perceive a strongly correlated region on a distribution of 2D scalar field.
This is related to the discrimination task concerning global structure on scalar fields\cite{reda2022rainbow}.
However, instead of discriminating global structure, we evaluate humans' abilities to identify the existence of strongly correlated regions, and their ability to accurately identify such regions.
Nevertheless, we hypothesize that the choice of color scale might still influence human effectiveness in this task.
Indeed for scalar field visualization, prior work\cite{mateevitsi2024science} has found that multi-hue color scale outperforms monochromatic luminance varying color scales, for helping participants extract and compare summary statistics in time-constrained settings.
Yet the task of identifying spatial correlation in scalar field distributions is more complex, and cognitively demanding, and thus the effects of color scale are less clear in this setting.


\subsection{Uncertainty visualization}

Uncertainty visualization has received considerable attention in recent years; we refer the reader to the following surveys~\cite{hullman2018pursuit,padilla2020uncertainty} for an overview of techniques and evaluation methods.
Uncertainty-based displays have, traditionally, focused on depicting a probability distribution governing a random variable, whether the distribution is provided in closed form or given as samples drawn from a distribution.
A common theme that drives uncertainty visualization design is to ensure the given display is actionable~\cite{boukhelifa2023visualization}, allowing one to make decisions on a particular task.
Standard methods for showing summary statistics~\cite{potter2010visualizing,correll2014error} have limitations in helping humans perform probabilistic reasoning for decision making.
This has motivated alternative methods for uncertainty displays, often depicting a continuous distribution as discrete events.

For instance, quantile dot plot displays~\cite{kay2016ish}, show evenly-spaced quantiles in a unit-based visualization, while HOPS~\cite{hullman2015hypothetical} depict an animation of draws from an underlying probability distribution, alongside other annotations necessary for a given task (e.g. a level of probability).
These designs have been shown to help humans more intuitively reason about probabilities through a frequency-based framing, e.g. ``a 20-out-of-100 chance for an event'', and thus quantities such as cumulative densities~\cite{ibrekk1987graphical} become easier to infer in a graphical depiction. 
When considering probability distributions that are more complex, e.g. not just a single random variable, then visualization design choices become more limited, as we must convey \emph{both} the underlying data, and its uncertainty.
Nevertheless, HOPs have shown quite versatile in supporting complex data governed by probability distributions, e.g. judging trends in times series~\cite{kale2018hypothetical}, performing topology-based tasks over networks~\cite{zhang2021visualizing}, as well as helping reason about missing data imputation~\cite{sarma2022evaluating}.

\subsection{Animation vs small multiples}
Due to the nature of temporal data or data distributions, static visualizations such as small multiples and dynamic visualization like animation have been widely used, and their effectiveness as has been extensively studied for a variety of analysis tasks \cite{robertson2008effectiveness, archambault2010animation, beecham2016map, boyandin2012qualitative, hosseinpour2024examining, farrugia2011effective, brehmer2019comparative}.
Robertson et al. \cite{robertson2008effectiveness} compared the effectiveness of small multiples and animation on trend visualization, and found that animation is fast and enjoyable to humans for presentation of trends. At the same time, animation is found to be less effective and accurate for the analysis of trend than small multiples. They point out one possible reason behind this: if humans are unfamiliar with the data, they do not know what part of the data will be salient. A consequence is that they might focus on the wrong region throughout the animation. We anticipate similar problems to arise in 2D scalar field distributions, a consequence of the rich and varying 2D information across draws.

Brehmer et al. \cite{brehmer2019comparative} further compared their effectiveness on trend visualizations, limited to mobile devices with small displays.
They advocated small multiples over animation since humans can identify the trend in less time with equal accuracy in most tasks, although they are slightly less confident. 
They found that small multiples can better help humans make detailed judgments on data such as angle or distance between trails, which must rely on memory or visual cues instead of eye movements during animation.
However, for tasks requiring comparison across a full 2D display, animation yield better accuracy than small multiples, since it benefits from shared coordinates rather than small multiples which use separate coordinates for individual plots.
For tasks related to identifying trajectory reversals, animation yield better performance than small multiples, which could be explained by Gestalt principle of 'common fate' in \cite{palmer1999vision}, that distractor items moving cohesively in one direction are likely to be perceptually grouped.

Archambault et al. \cite{archambault2010animation} compared the effectiveness of animation and small multiples visualization of dynamic graphs, on helping human gain understandings about the local(e.g. node degree changes; node appearance) and global(e.g. number of nodes; shortest path) properties of graph evolution over time.
They found that small multiples seems to perform better on topology-based tasks.
This might be due to humans having sufficient time to read topological information in all time slices at the same time.
On the other hand, animation has significantly higher accuracy than small multiples when the inference tasks are related to appearance of nodes or edges in the dynamic graph.
This indicates that the animated transitions between timeslices of dynamic graph is helpful in inference tasks made along a temporal axis.




%% file: sections/background.tex
In this section we discuss visualization design choices, those that support making inferences on spatial correlation for 2D scalar field distributions.
We conclude with hypotheses about their effectiveness, which serves as the basis for our study.

\subsection{Background}

\begin{figure}[!t]
    \centering
    \includegraphics[width=.49\textwidth]
    {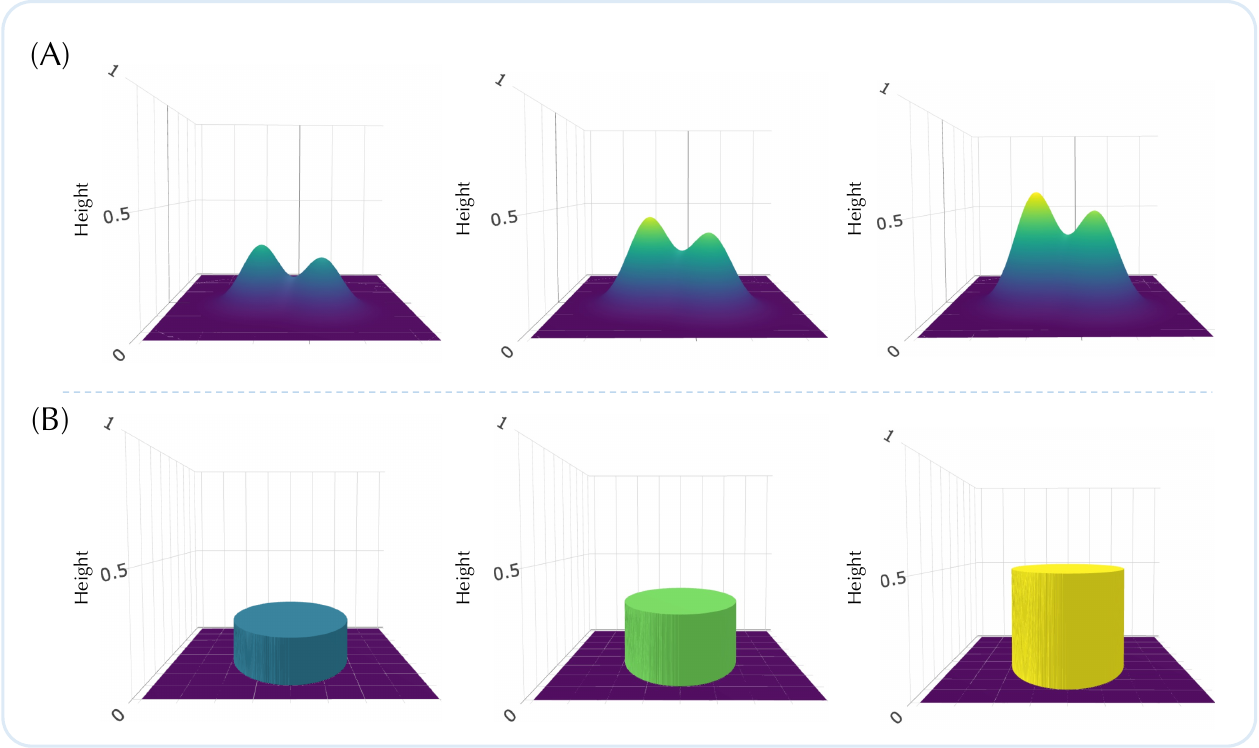}
    \caption{The strongly correlated regions on the 2D height fields move up and down together in both (A) and (B). While (A) each location inside the region has its own height, which forms a complex 2D surface.
    in (B) all locations inside that region share the same height, forming a flat surface. 
    }
    \label{fig:height_field}
\end{figure}

Correlation is a statistical measure that describes the extent to which two random variables change together. 
For example, given two locations on a 2D height field, we may view their heights as samples from two underlying random variables, governed by a joint probability distribution.
If the two locations have zero correlations, then they will change arbitrarily across draws from the distribution.
On the contrary, if two random variables have a high positive correlation, then their heights are very likely to go up and down simultaneously for draws taken from the joint distribution.
However, the visual patterns that result from high correlation can vary.
A given region for which all pairs of points have high correlation might manifest as values that are roughly the same within draws, but vary between draws.
This implies a flat surface on a 2D height field moving up and down together as shown in Fig. \ref{fig:height_field}(B).
Alternatively, the region might not be of constant height, but rather vary in value, which means a 2D surface with a special spatial structure that, nevertheless, induces a consistent change across draws, shown in Fig. \ref{fig:height_field}(A).
In this paper, we are interested in the former, whereby a strongly correlated region gives similar values across draws.
More specifically, we want to investigate how human perceive this kind of strongly correlated region on a distribution of 2D scalar field, through different visualization designs.

As previously discussed, numerous methods exist for extracting and visualizing strongly correlated regions (c.f. Sec.~\ref{sec:rw_corr}).
Alternatively, we may opt to show summary statistics concerning the data, e.g. the mean field, as well as per-location variance.
However we eschew these visualization forms, and instead focus on simplicity in our study: directly show the data, as-is, via color-mapped scalar fields.
If users can accurately make correlation judgments with such designs, this indicates broader application of simple visualizations for analyzing complex data sources, calling into question the need for elaborate techniques.

Towards this end, we hypothesize three main data characteristics of scalar field distributions that will influence human perception of spatially correlated regions.
The first is the correlation strength: the higher the value, the more consistently humans will identify a region as being strongly correlated.
Secondly, we hypothesize that the size of the region is important.
The larger the region, the easier it will be for humans to recognize correlation.
The last factor is the discriminability of the correlated region, namely the level of contrast with the background.
The higher the discriminability, the easier it should be for humans to recognize correlation.

As for the visualization design, we are interested in comparing two fundamental visualization types: small multiples and animation. Small multiples
allows users to perform a detailed analysis on individual samples of the distribution.
The design helps humans store detailed information about individual draws, e.g. the existence or size of the region with identical values, through visualization of individual plots, rather than having humans store such information in memory \cite{todd2004capacity}.
In addition, a small multiples design leverages the human visual system’s ability to ensemble and contrast across different static representation \cite{tufte1983visual} at the same time.
This helps humans quickly identify commonalities and differences across multiple draws, e.g. the maximal region that consistently has identical values across draws.
However, small multiples has an inherent spatial alignment issue which may introduce error when humans are trying to compare the locations of regions across draws.

Regarding animation, as studied in many previous work about map visualization and uncertainty visualization, animation of color maps is a common way to convey draws from a 2D distribution.
We consider animation of color-mapped 2D scalar field distribution as an instance of hypothetical outcome plots \cite{hullman2015hypothetical} (HOP), with each frame representing a single sampled scalar field from the distribution. 
Animation provides for a direct comparison across draws, \emph{in situ}, without spatial alignment issues like small multiples, which may improve the accuracy of humans' perception about the location and size of the spatial correlated region. 
However, animation might lead humans astray, as it might be unclear what to attend to in completing the task.
Specifically, when human are looking for a correlated region, they may be attracted to multiple regions due to the rich information stored in scalar fields.
This is different from 1D distributions\cite{hullman2015hypothetical} where human have a clear idea about where to look during the animation.
Since we do not allow participants to replay the animation, they may choose a specific region which is likely to be the strongly correlated region after the first few frames, and focus their attentions on that region in subsequent frames to verify their assumptions.
If humans attend to the wrong region at first, they will very likely obtain the wrong impression after watching the animation.
Moreover, animation may provide insufficient time for humans to analyze and compare individual draws in detail, which might make it hard to even recognize the target region of high correlation. This is exacerbated when the correlation strength is low, the area of the region is small, or the discriminability of the correlated region is low.


In addition to visualization types, we are also interested in how the color encoding of 2D scalar fields influences human judgment. Based on previous work on the effects of color categorization on human performance in discrimination tasks, we hypothesize that multi-hue color scales will help humans better discriminate the target region of similar colors from the background, characterized as independence across draws. Since multi-hue color scales tend to provide wider span of hues for changes and variations, we want to investigate whether humans can ensemble the information provided by multi-hue color scales across multiple draws, without exceeding their memory limit or resulting in cognitive overload.


\subsection{Hypotheses}
Based on the above discussion, we propose three hypotheses on how different visualization designs will influence human perception:


\textbf{\underline{(H1)}}  
Small multiples will give an improvement over animation in helping humans recognize strongly correlated regions in scalar fields.

\textbf{\underline{(H2)}}  
Small multiples will introduce higher error in humans describing the strongly correlated region, in comparison to animation.

\textbf{\underline{(H3)}}  
Multi-hue luminance-varying color scale will make it easier for humans to identify the existence of a spatially correlated region.

These hypotheses inform the factors that underlie our experimental design, which we turn to next.




%% file: sections/study_design.tex
\subsection{Data generation}

In generating stimuli for our study, we wish to control for (1) \textbf{correlation strength}, (2) the \textbf{size} of the target region, and (3) \textbf{discriminability} of the target region from the background.
Moreover, we wish to vary (1) the shape of the target region, and (2) the location of the region, to mitigate effects of recall in what participants should be looking for.
Towards theses goals, we represent a distribution of scalar fields as a multivariate Gaussian.
A scalar field is represented as a discrete set of samples over a 2D domain, in practice a square regular grid with fixed spatial resolution $s \times s$.
We may thus flatten this 2D grid of values to a vector, $\mathbf{z} \in \mathbb{R}^N$ with $N = s \cdot s$, and it is this vector which is governed by a multivariate Gaussian.
The mean parameter is fixed to zero, while the covariance matrix, which we denote by $\boldsymbol{\Sigma}$ reflects all data characteristics:
\begin{equation}
    \Sigma_{i,j} = \exp\left(-\lVert \mathbf{x}_i - \mathbf{x}_j \rVert^2 / \sigma_l^2\right) + v_i \cdot v_j,
    \label{eq:cov}
\end{equation}
for row $i$ and column $j$, corresponding to positions $\mathbf{x}_i, \mathbf{x}_j \in \mathbb{R}^2$ in the domain of the field, respectively.
We break down the above equation in the following.

\textbf{Background model.} The term $\exp(-\lVert \mathbf{x}_i - \mathbf{x}_j \rVert^2 / \sigma_l^2)$ is intended to represent random background noise when taking draws from the Gaussian.
The term $\sigma_l^2$ is the length scale, controlling for the similarity between pairs of points.
As $\sigma_l \rightarrow 0$, this gives a diagonal covariance (excluding the foreground), and thus independence across positions.
Otherwise, the value of $\sigma_l$ controls the level of dependence for two positions $\mathbf{x}_i$ and $\mathbf{x}_j$ in the domain.
If their Euclidean distance is sufficiently large, relative to $\sigma_l$, then they will take on similar values when drawing samples.

\textbf{Foreground model.} We construct a vector $\mathbf{v} \in \mathbb{R}^N$ to determine the target region, namely, a spatial region in which all pairs of points have high covariance.
Specifically, for a given index $i$ with associated 2D position $\mathbf{x}_i$, each entry of $\mathbf{v}$ is:
\begin{equation}
v_i = \sigma^2_v \exp\left(-(\mathbf{x}_i - \mathbf{s})^T \mathbf{M} (\mathbf{x}_i - \mathbf{s})\right).
\label{eq:v}
\end{equation}
When formed as an outer product $\mathbf{v} \mathbf{v}^T$, and added to the covariance (c.f. Eq.~\ref{eq:cov}), draws from the distribution give the appearance of an ellipse at a specified region.
Specifically, the vector $\mathbf{s} \in \mathbb{R}^2$ is the center of the region, while the matrix $\mathbf{M} \in \mathbb{R}^{2 \times 2}, \mathbf{M} := \mathbf{Q} \boldsymbol{\Lambda}^2 \mathbf{Q}^T$ controls for the orientation (rotation matrix $\mathbf{Q}$), as well as the lengths of the semi major \& minor axes (entries on diagonal matrix $\boldsymbol{\Lambda}^2$).
If the product $v_i \cdot v_j$ is large, then their corresponding points $\mathbf{x}_i$ and $\mathbf{x}_j$ are more likely to be in the interior of the ellipse.
The parameter $\sigma^2_v$ controls for the influence of the foreground, relative to the background.

Once we have constructed the covariance matrix, we then derive a correlation matrix via $\bar{\boldsymbol{\Sigma}}$, and arrive at our scalar field distribution $p(\mathbf{v}) = \mathcal{N}(\mathbf{v} | \mathbf{0}, \bar{\boldsymbol{\Sigma}})$.
We use the correlation, rather than covariance, to ensure that the scale of the values across draws is consistent.
The visualization designs for our study display samples drawn from this distribution.
We control for the factors of our data via the following:

\textbf{Correlation strength.} The parameter $\sigma^2_v$ represents the strength of correlation for the target region.
For a sufficiently large value of $\sigma^2_v$, all pairs of points within the ellipse of our foreground model will have a large covariance value, in comparison to pairs of points for which one point is inside of the ellipse, and the other is outside of the ellipse.
On the other hand, we wish to ensure that there is a smooth transition between foreground and background, thus making less obvious the appearance of the ellipse.
To achieve this, we adjust the diagonal matrix $\boldsymbol{\Lambda}^2$ such that for points $\mathbf{x}_i$ at the boundary of the ellipse, their corresponding values $v_i$ are sufficiently small, approximately $0.1$.

\textbf{Size (Shape length).} The diagonal matrix $\boldsymbol{\Lambda}^2$ gives control on the size of the target region, e.g. the ellipse shape.
Specifically, we associate size with the length of the minor axis of the ellipse -- we denote this by $\lambda_s = \sqrt{\Lambda_{11}^2}$.
The major axis is assigned a random value, at most twice the length of the minor axis.
This is done to introduce randomness in the presented shapes, and in turn suppress participants from memorizing the shape that they should be looking for.

\textbf{Discriminability.} The length scale parameter $\sigma_l$, and the size parameter $\lambda_s$, jointly represent the discriminability of the target region.
We measure discriminability via the following ratio:
\begin{equation}
a(\sigma_l, \lambda_s) = \max(\sigma_l / \lambda_s , \lambda_s / \sigma_l).
\label{eq:ratio}
\end{equation}
When $a$  is close to 1, the spatial scale of the background model is comparable to that of the foreground target region.
The target region will nevertheless remain distinct, since all pairs of points within the ellipse shape will take on similar values within a single draw (c.f. Fig.~\ref{fig:height_field} top row), in contrast with the background model giving noise of a limited set of spatial frequencies (c.f. Fig.~\ref{fig:height_field} bottom row).

\begin{figure}[!t]
    \centering
    \includegraphics[width=.49\textwidth]
    {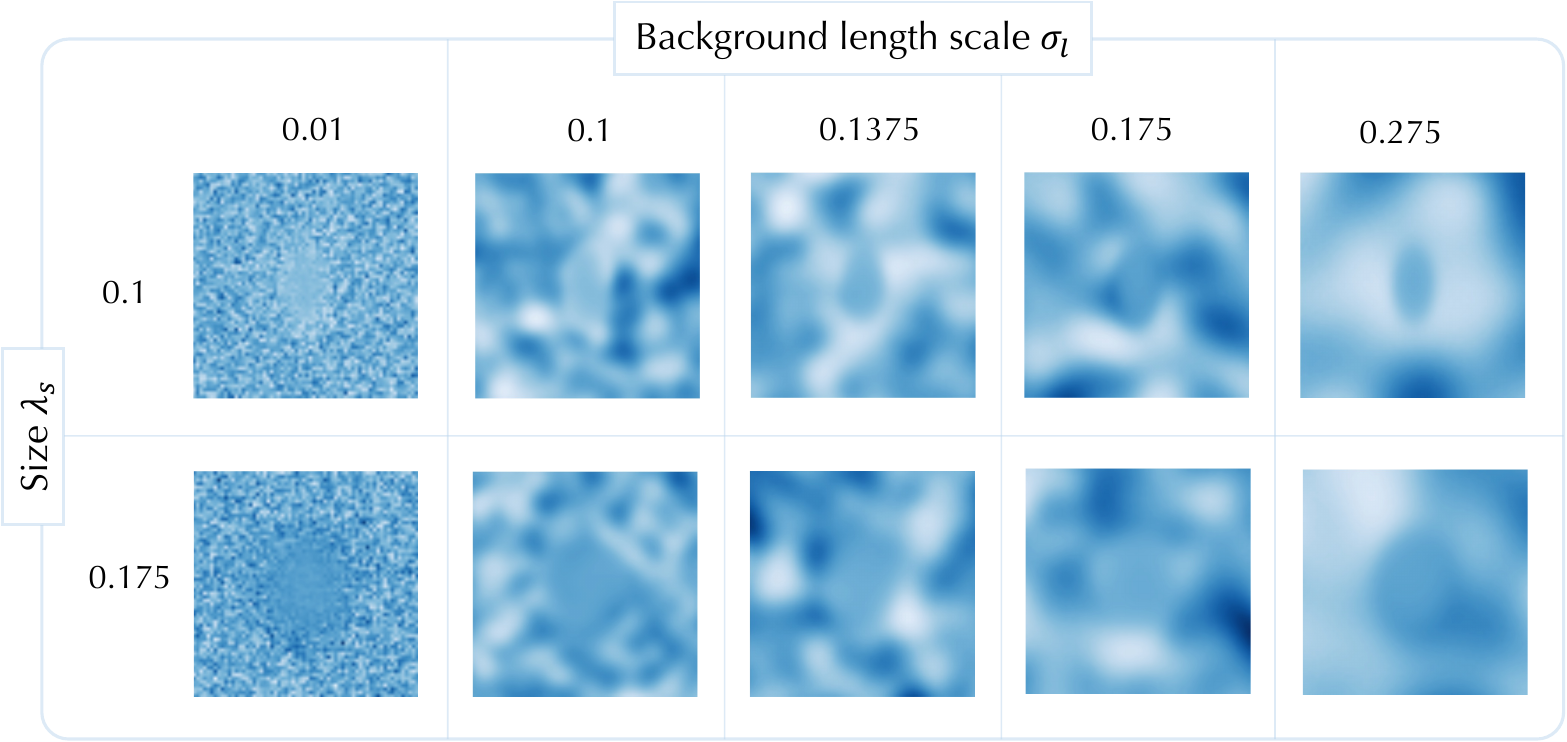}
    \caption{We show draws from different 2D scalar field distributions, giving different levels of discriminability for the target correlated region, relative to the background. Shown are 5 levels of the background length scale [0.01, 0.1, 0.1375, 0.175, 0.275], and 2 levels for size of the target region [0.1, 0.175]. When both of these are equal (0.1 or 0.175), then the level of discriminability is smallest, making the target region challenging to recognize. In contrast, a small length scale (0.01) gives rise to high-frequency noise, and thus eases recognition of the target region.}
    \label{fig:data}
\end{figure}

To better illustrate the factors of size and discriminability, in Fig.~\ref{fig:data} we show draws from different distributions that vary by background length scale $\sigma_l$ (columns) and target region size $\lambda_s$ (rows), holding fixed correlation strength.
For a sufficiently small $\sigma_l$, we find that the background model gives rise to high-frequency noise, for which we expect the target region should be easy to recognize.
On the other hand, when $\sigma_l$ and $\lambda_s$ are equal, e.g. either 0.1 or 0.175, then the frequency content of the noise will be compatible with the ellipse shape, thus making it more difficult to discriminate the target region.

Based on pilot studies, we decided that 2 levels for size $\lambda_s$ (0.1, 0.175), and 5 levels of background length scale $\sigma_l$ (0.01, 0.1, 0.1375, 0.175, 0.275) allow us to study a variety of distributions that vary in difficulty level.
To mitigate memorization of the target region across trials, we randomize the center position of the ellipse ($\mathbf{s}$) and the orientation ($\mathbf{Q}$).
The correlation strength is set to 4 levels (1.5, 2, 2.5, 3).

\subsection{Visual encodings of scalar fields}

For the color encoding of scalar field distributions, we choose ``Viridis'' and ``Blues'' as representatives for multi-hue and single-hue luminance-varying color scales, respectively.
Given a range of values, the Blues color scale will vary the intensity of blue from light blue for a small value, to dark navy for a large value.
Viridis spans a range of hues, transitioning smoothly from dark purple and blue at the low end, through green and yellow, to bright yellow at the high end. 

These choices are motivated by prior work that has studied the use of color scales that vary in hue for encoding quantitative data. A conventional color design for quantitative data is one that strictly varies in luminance, in particular for resolving details and emphasizing contrast~\cite{ware2019information}, both important aspects of our designed task. The Blues and Viridis color scales are both designed to be perceptually uniform in luminance, and so one might not expect variations in hue to offer any additional cues to the user. Yet prior work has shown the effects of multi-hue color scales on color~\cite{liu2018somewhere} and pattern~\cite{reda2020rainbows} discrimination. Our setting is distinct in that users must aggregate multiple views to make a judgment, and so it is unclear whether these prior studies are entirely predictive of the effects of multi-hue versus single-hue color scales.

\subsection{Design of tasks}
\label{subsec:dot}

Our study aims to test how well participants can recognize a spatially correlated region.
Moreover, if they do recognize the region, we further aim to measure the accuracy of their description of the region.

\textbf{Task 1: recognition.} We have designed a task to let participants provide an estimation of \emph{how frequently} they notice the spatially correlated region.
For animation, we let participants provide their estimations of frequency as a percentage from 0\% to 100\%.
Although prior work has shown the merits of \emph{counting}~\cite{hullman2015hypothetical}, e.g. 10-in-100 times, users in our pilot studies found it more natural to express frequency as a percentage.
For small multiples, we let participants provide a count of how many images, within the display, contain a consistent target region that is distinct from the background.
This is contrast with the percentage framing of animation, as the number of images one can show in a small multiples display is limited, and thus counting the number of images is feasible to perform, and report.
Given that each trial contains a target region of correlation, users might be tempted to always give an answer of 100\%. Yet the lowest level of correlation and discriminability gives visualized fields for which the target region is very difficult to observe. Thus we decided that these trials would inhibit such pathological cases, and from the collected data, no participants answered with exclusive 100\% frequencies.

\textbf{Task 2: description.} In the task of recognition, although participants might be confident in reporting a strongly correlated region, it is possible that what they perceive is incorrect, e.g. the actual target region differs from what they perceive.
Thus, for our second task we wish to measure the accuracy of participants in describing the region that they perceive.
For each trial in our study, upon conclusion of the visual display, we show a blank canvas in place of the visualization.
We then let participants brush, on this blank canvas, the target region of high spatial correlation.
A single click of the brush gives a small, 5 pixel-radius region of values varying between 0 and 1, where the values smoothly decay away from the center according to a Gaussian function.
If a participant does not manage to find any spatial correlated region during their observation, they may specify that they ``have not seen any pattern'', without providing their frequency estimations or brush.

To measure the accuracy of the user's brush, we wish to quantify the spatial alignment between the user-specified region, and the ground-truth region.
We consider two metrics for alignment: the intersection-over-union metric (IoU), and the parameters corresponding to an affine transformation designed to explicitly align the regions.

\textbf{IoU.} To compute IoU, we first need to convert both the user's brushed region, and the target correlated region, into binary masks, where 1 indicates foreground, and 0 indicates background.
As the user's brushed region is continuous, we binarize it by converting all values to 1 whose brushed values exceed 0.2, and 0 otherwise.
For the target region, we find that a size-dependent threshold is necessary to obtain masks that are consistent with the color maps.
Specifically, we convert all values of Eq.~\ref{eq:v} to 1 if they exceed 0.7 for size of 0.1, and exceed 0.99 for size of 0.175, and 0 otherwise.
We then take the user's binarized region, denoted $H$, and the binarized ground truth, denoted $G$, and compute:
\begin{equation}
\label{eq:iou_binary}
\text{IoU} = \frac{\sum \left(\text{H} \cdot \text{G}  \right)}{\sum \text{H} + \sum \text{G} - \sum \left(\text{H} \cdot \text{G} \right)},
\end{equation}
where the sum is over all locations in the image.

\textbf{Affine transformation.} As previously discussed in Sec.~\ref{sec:background_hypothesis}, we anticipate that a small multiples display will result in higher error for participants in describing the region, due to issues of spatial misalignment.
This should manifest as higher IoU, despite the fact that the participant might have accurately recognized the target region.
To counter issues of spatial misalignment, we explicitly align the user's brushed region with the ground-truth.
Specifically, for the binarized target region, we find the affine transformation that best matches the user's binarized brushed region.
We decompose an affine transformation into (1) a translation, (2) an isotropic scaling, and (3) a rotation.
We perform a dense grid search over these three parameters, and select the combination of parameters whose IoU with the brushed region is highest.
We use these parameters as a way to better explain the cause for high IoU, e.g. whether a user is accurately perceiving the scale of the object, and only incurring error in its location.

\subsection{Interface Design}
\label{sec:interface}

For the animation and small multiples conditions, there are a number of important parameters that need to be set, namely the duration of time for which the visualization is shown, the number of fields to display, and specific to animation, the frame rate.
We adopt the following principle in choosing parameters: we aim to maximize the respective strengths of the different visualization conditions, given a limited time window to complete the task.
To this end, for the animation condition, we ultimately settled on a frame rate of 6 frames per second (fps), and animation time of 10 seconds.
These parameters were determined via pilot studies.
Specifically, we initially chose a frame rate of 10 fps, but participants reported the animation was too fast to make accurate judgments, a consequence of the lack of ordering, and thus lack of continuity, across draws.
Moreover, we initially set the animation time to be 20 seconds, yet post-study, participants reported that they were able to complete the task in a shorter amount of time.
For each trial we randomly sampled 30 scalar fields from a given distribution. We replicated each sample twice and shuffled the sequence of all samples. The first and last frame of the animation is set to a blank canvas.
Once users click on the Play button, they cannot pause the animation or slide back and forth.
Participants use a slider to report the frequency (0 - 100\%) with which they observed the region of spatial correlation.
Once the slider is set, participants then have the opportunity to brush over the blank canvas to specify where they believe the correlated region lies.
If participants do not observe such a region, they do not need to brush, and instead specify, via a button, they ``have not seen any spatial correlated region''.

For the small multiples condition, we set the time duration to also be 10 seconds, and thus participants across conditions spend equal time on each trial.
As in the animation condition, we show a blank canvas at the beginning and end of a trial, specifically in the top-left portion of the plot region, and users must complete the same tasks at the conclusion of 10 seconds.
However, different from animation, we decide to show 10 scalar fields, arranged in a $2 \times 5$ grid, randomized to inhibit participant dependency on order.
This choice has the limitation that data is not equivalent between conditions.
If we ensure data is the same across conditions, then in choosing a large number of fields (e.g. 30), this plays to the strength of animation where we can easily show all such data. However for small multiples this leads to an excessive number of views to analyze and reconcile in a time-constrained setting, and recent work gives evidence that large view count can impair performance~\cite{hosseinpour2024examining}. Conversely, a small number of fields (e.g. 10) makes the small multiples design more manageable, yet we have few frames to display for animation. In assigning a larger number of fields to animation than small multiples, this ensures we are best utilizing both designs.

\subsection{Study Procedures}

We conducted a 2 (visualization type, between) × 2 (color scale, between) x 4 (correlation strength, within) x 5 (background length scale, within)  × 2 (size, within) mixed-factors design for our study.
We treat visualization and color scales as between-subjects factors to ensure consistent responses, and prevent comparative judgments across different visual encodings. 
Participants complete a total of 40 trials, where we randomly generate a distribution of 2D scalar field based on each combination of (correlation strength, background length scale, shape length) for each trial.
The position and orientation of the ellipse for the target function of each 2D scalar field distribution is randomly sampled on the 2D domain.
We shuffled the sequence of trials for each participant in order to inhibit user dependencies and bias on the sequence of trials.

Participants were recruited through Prolific, and completed the study remotely via a hosted website.
We recruited 88 participants in total, and assigned 22 participants to each combination of between subject factors.
This study has been approved by the Institutional Review Board (IRB) at the authors’ institution.
We provide instructions and qualification test before the main study.
In the instructions, we describe the task that participants are expected to complete, and provide participants with a qualification test across 2 trials that we determined to contain regions of high spatial correlation.
Participants must successfully complete the qualification to take part in the study.
Pilot studies suggested that our full study – instructions, qualification, and 40 trials – required approximately half an hour to complete. 

%% file: sections/result.tex
In this section, we present our findings from two aspects.
First, we study how visualization design, and data characteristics of the scalar field distribution, influence the recognition of strongly correlated regions.
Secondly, we study how such factors influence participant accuracy in describing these regions.
For the convenience of notation, in the following sections, we will denote visualization type as \emph{vis}, color scale as \emph{color}, correlation strength as \emph{corr}, length of the target region, namely semi-minor axis of ellipse, as \emph{shape\_l}, and background length scale as \emph{background\_l}.

\subsection{Correlation recognition}
\label{analysis:freq}

\subsubsection{Model Design}
Since the estimation of frequency participants provided in each trial is a continuous variable ranging from 0 to 1, we employed a generalized linear mixed-effect model with beta regression to study what factors – and factor combinations – are predictive of our response variable.
Beta regression provides a flexible way to model a bounded continuous variable representing a proportion ranging from 0 to 1 (excluding 0 and 1).
We transformed our response variable of frequency based via:
\begin{equation}
freq\_adjust = \frac{freq \times (n - 1) + 0.5}{n},
\label{eq:beta_regression}
\end{equation}
in order to ensure the dependent variable excludes 0 and 1.

We consider all possible interactions among all factors discussed in Section \ref{sec:study_design} in our initial model.
We iteratively remove interactions via  backward model selection process, where we use the Bayesian information criteria as the basis for removal, favoring simple models that still fit well to observations. In the end, we obtain:
\begin{align*}
\text{freq\_adjust} &\sim \text{Beta}(\mu, \phi) \\
\text{logit}(\mu) &= \text{vis} \times \text{shape\_l} + \text{corr} + \text{color} : \text{corr} \\
&\quad + \text{vis} : \text{color} : \text{shape\_l} \\
&\quad + \text{shape\_l} : \text{background\_l} \\
&\quad + \text{shape\_l} : \text{background\_l} : \text{vis}\\
&\quad + \text{shape\_l} : \text{background\_l} : \text{corr}\\
&\quad + (1 \mid \text{user})
\end{align*}
In what follows, we list relevant effects for individual variable found through our fitted model, as well as interactions between variables.

\subsubsection{Results}

\begin{figure}[!t]
    \centering
    \includegraphics[width=.5\textwidth]{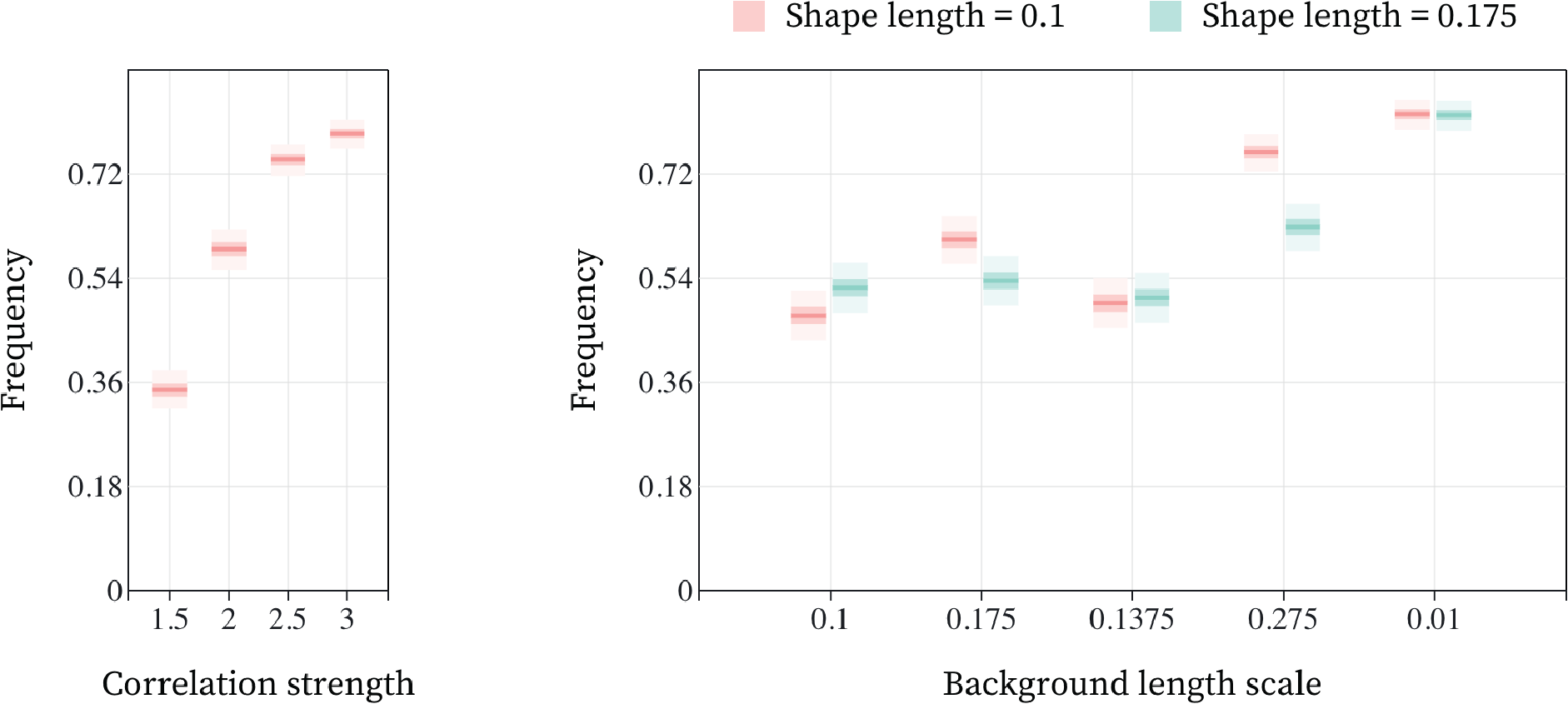}
\caption{We summarize results for recognition of correlation.  A higher correlation strength, or level of discriminability, helps participants recognize the existence of a strongly correlated region, regardless of visualization design.}
    \label{fig:freq_noise_rl}
\end{figure}

\textcolor{midblue}{[shape\_l; corr; shape\_l $\times$ background\_l]} We first start with data factors. As shown in Figure \ref{fig:freq_noise_rl}, we found that the correlation strength ($\chi^2 = 1062.0631, Df = 3, p < \num{e-5}$), shape length ($\chi^2 = 6.5606, Df = 1, p = 0.010426$) and discriminability of the strongly correlated region ($\chi^2 = 732.7736, Df = 8, p < \num{e-5}$) 
all have significant effects on self-reported frequency.
The higher the correlation strength, the more likely participants can recognize the strongly correlated region.
Moreover, the larger the length of the target region, or higher level of discriminability, the easier it is for participants to recognize the target region.
This is independent of visualization design.

\begin{figure}[!t]
    \centering
    \includegraphics[width=.49\textwidth]{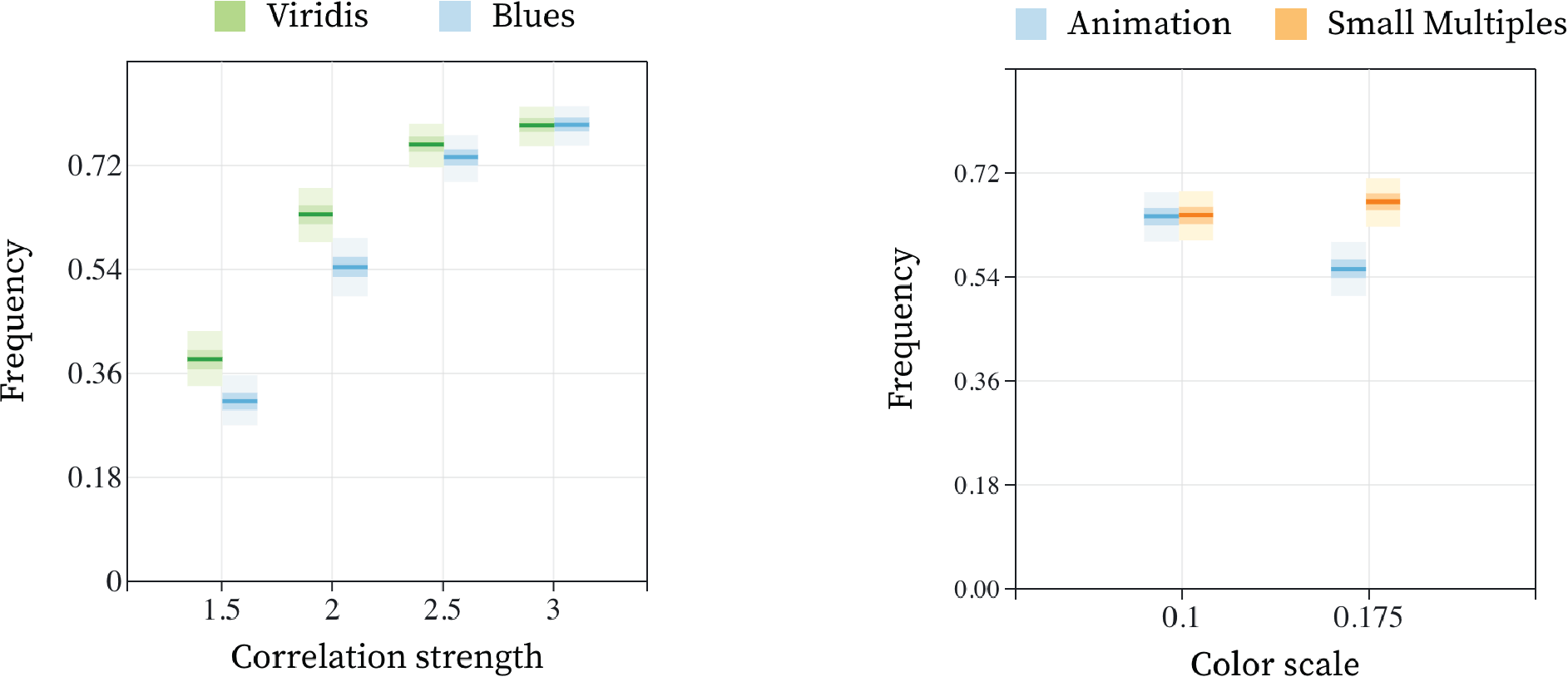}
    \caption{Our study reveals that recognition of correlation across different color scales is impacted by correlation strength (left) -- the weaker the correlation, the additional details in a multi-hue color scale (Viridis) aids in recognition. Moreover, we find that small multiples leads to higher recognition, but only when the shape length is large, suggesting that analyzing a set of juxtaposed views benefits pattern recognition.
   }
    \label{fig:freq_color_vis}
\end{figure}

\textcolor{midblue}{[color $\times$ corr]} In considering the color scale, we find an interaction with correlation strength.
More specifically, as shown in Fig. \ref{fig:freq_color_vis} (left) participants tend to recognize the existence of the target region ($\chi^2 = 14.4865, Df = 4, p = 0.005894$) more frequently when the scalar field distribution is visualized using Viridis color scale, compared to Blues color scale, but only when the correlation strength is relatively low, such as 1.5 and 2.

\textcolor{midblue}{[vis $\times$ shape\_l]} Inspecting visualization type, we find a significant interaction with shape length ($\chi^2 = 33.5110, Df = 1, p < \num{e-5}$).
Using estimated marginal means, we find no difference between the effects of small multiples or animation on recognition when the shape length is relatively small (0.1), as shown in Fig. \ref{fig:freq_color_vis} (right). However, when the shape length is relatively large (0.175), small multiples has a significantly ($odds.ratio = 0.61645, SE = 0.0831, z.ratio = -3.59, p =0.00033$) higher effect on the response variable than animation.

\begin{figure}[!t]
    \centering
    \includegraphics[width=.5\textwidth]{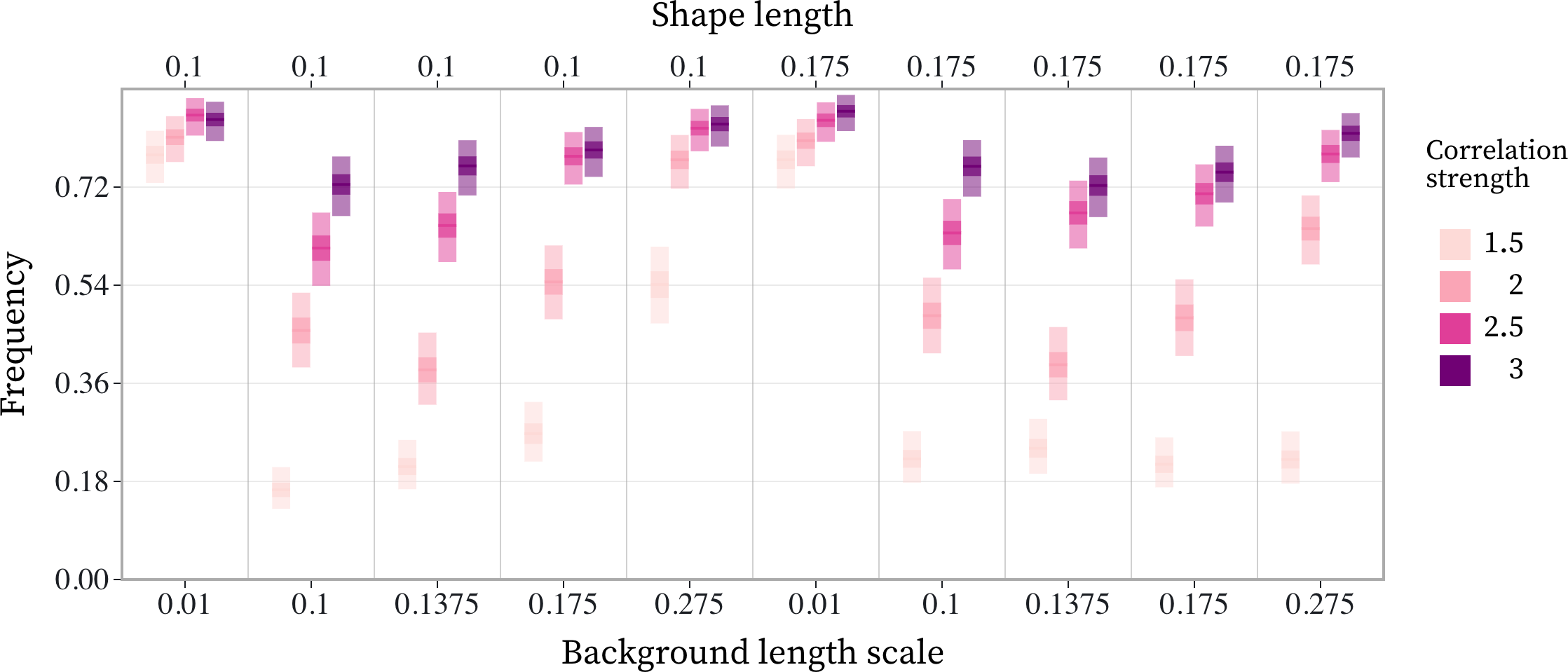}
    \caption{We show a summary of 3-way interactions between correlation strength, length of target region, and background length scale. We observe the strongest effect of correlation strength when the ratio between shape length and background length scale (c.f. Eq.~\ref{eq:ratio}) is closest to 1, representative of a small level of discriminability.}
    \label{fig:freq_corr_size_back}
\end{figure}

\textcolor{midblue}{[corr $\times$ shape\_l $\times$ background\_l]} We additionally find interactions between correlation strength, shape length, and background length scale ($\chi^2 = 252.3033, Df = 27, p < \num{e-5}$).
We find that correlation strength plays a stronger role when the discriminability level is low, as shown in Fig.~\ref{fig:freq_corr_size_back}
Namely, when the strongly correlated region is less distinguishable from background, the consistency in the perceived pattern (high correlation strength) becomes more relevant in recognition.

\textcolor{midblue}{[vis $\times$ shape\_l $\times$ background\_l]} We further explored how discriminability level impacts recognition, and found a significant interaction between visualization type, shape length, and background length scale ($\chi^2 = 42.1022, Df = 8, p < \num{e-5}$).
We analyze the result using estimated marginal means, and find that the small multiples design gives higher self-reported frequencies than animation.
This is most prominent when the discriminability level is highest, or when the length of the ellipse, and spatial scale of background noise, are similar (c.f. Eq.~\ref{eq:ratio}).

\textbf{Discussion:} our results highlight that, under certain conditions, small multiples leads to higher recognition compared to animation (\textbf{H1}).
Specifically, the size of the ellipse becomes important, suggesting that a large region of correlation might be difficult to rapidly perceive as part of an animation, whereas with small multiples, participants have more time to analyze, and integrate, a set of views to help identify the region.
Moreover, our results indicate that the Viridis color scale can lead to higher recognition (\textbf{H3}), but only for the more challenging scenario of lower correlation strength.
A low correlation strength can lead to low contrast between foreground and background across draws from the distribution.
A multi-hue color scale can offer more detail, e.g. color variation, gradient, and thus is likely supportive in recognizing the target region, consistent with prior work~\cite{reda2020rainbows,reda2022rainbow}.

\subsection{Describing the correlated region -- IoU}
\label{analysis:iou}

\subsubsection{Model Design}
We employed a linear mixed-effect regression model to study what factors – and factor combinations – are predictive of spatial alignment, as measured through IoU. Similarly to Sec.~\ref{analysis:freq}, we considered all possible interactions among factors in our initial model, and through refinement we arrive at the following for our model:

\begin{align*}
\text{iou} &\sim \text{N}(\mu, \sigma^2) \\
\mu &= \text{vis} \times \text{shape\_l} + \text{corr} \times \text{shape\_l} \\
&\quad + \text{shape\_l} : \text{background\_l} \\
&\quad + \text{shape\_l} : \text{background\_l} : \text{corr} \\
&\quad + (1 \mid \text{user})
\end{align*}

\subsubsection{Results}


\textcolor{midblue}{[vis; vis $\times$ shape\_l]} We first explore how the visualization designs impact accuracy.
We find that the accuracy for animation is significantly ($\chi^2 = 16.6087, Df = 1, p = \num{4.594e-05}$) more accurate than small multiples visualization.
This verifies our hypothesis (\textbf{H2}) regarding the inherent spatial alignment issues of small multiples. 
We further find interactions with visualization and shape length ($\chi^2 = 7.6138, Df = 1, p = 0.0058$), in particular, we see a boost in accuracy for small multiples when the shape is larger ($estimate = -0.0614, SE = 0.00822, Df = 2877,	t.ratio = -7.466, p <0.0001$); this effect is not found in animation.
This indicates that spatial alignment is most pronounced when the ground-truth region requires a high level of precision to accurately describe.

\begin{figure}[!t]
    \centering
    \includegraphics[width=.5\textwidth]
    {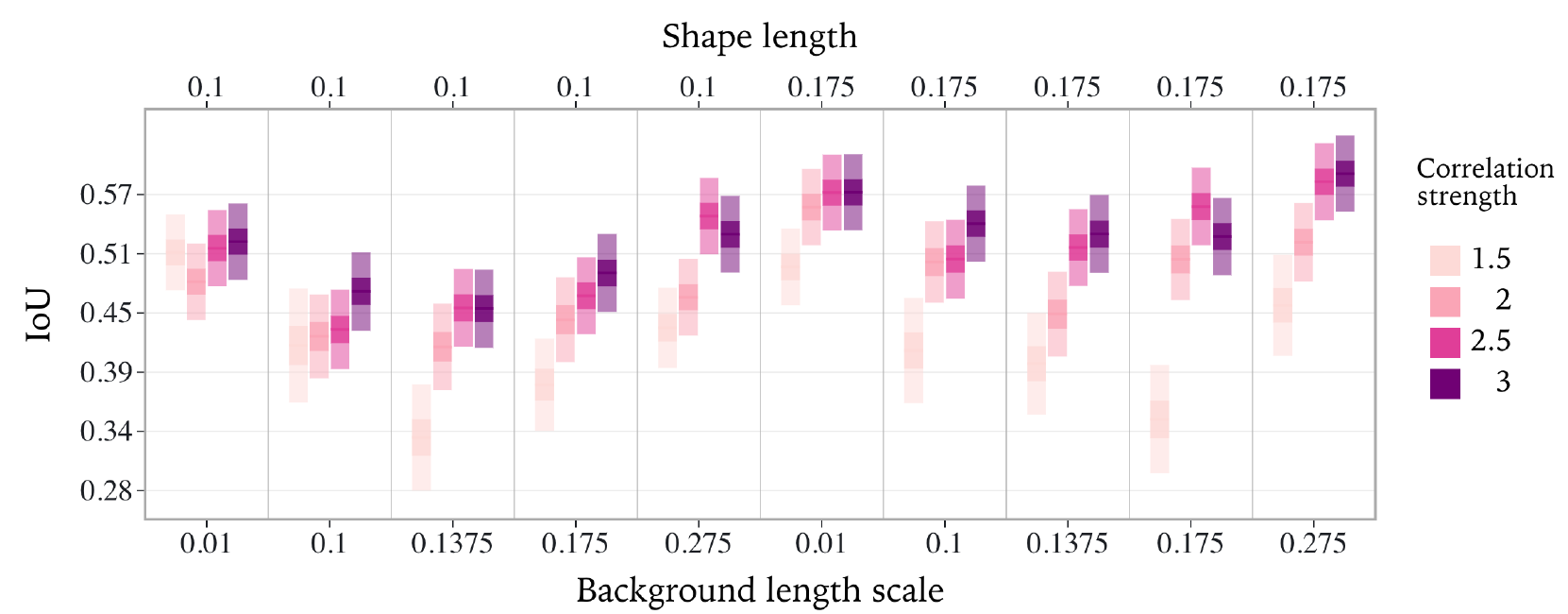}
    \caption{We show the accuracy of participants in describing the region, measured in terms of IoU, summarized over all visualization conditions. We find (1) participants are less precise when the target region is small, and that (2) discriminability impacts performance, especially when the correlation strength is low.}
    \label{fig:iou_noise_rl}
\end{figure}

\textcolor{midblue}{[shape\_l $\times$ background\_l $\times$ corr]} Similar to the recognition scenario, for IoU-based alignment, we find a significant interaction between shape length, background length scale, and correlation strength, please see Fig.~\ref{fig:iou_noise_rl} for a summary.
In particular, for a low correlation strength, participant accuracy is reduced ($\chi^2 = 44.9969, Df = 24, p = 0.005830$).
We find that the shape length impacts participant accuracy -- participants are less precise in their brush for smaller target regions.
Moreover, the correlation strength impacts ($\chi^2 = 44.9969, Df = 24, p = 0.005830$) participant accuracy, in particular, when the discriminability level is small.
For example, when both shape length and background length scale are 0.175, participants have significantly lower accuracy ($estimate = -0.206, SE = 0.0305, Df = 2841.67,	t.ratio = -6.7486, p < \num{e-05}$) 
when correlation strength is relatively low, e.g. 1.5, compared to a high correlation strength, e.g. 3.5.
In other words, when the strongly correlated region is not distinguishable compared to its background, correlation strength starts to play an increasingly important role in participant accuracy.

\textbf{Discussion:} though we observe a reduction in performance for the small multiples condition, this needs to be understood in the context of IoU, as just one way to measure accuracy via position-wise alignment.
As such this might not represent a user's perception of the target region.
In observing a small multiples arrangement, a participant might form a good conception of the region, yet the necessity to aggregate \emph{multiple} views can lead to imprecision in specifying the region within just a \emph{single} view.
The consequences of this imprecision are less clear, as the specified region could be: (1) off by a translation, (2) equivalent to the target region upon application of a uniform scaling, or (3) equivalent, under a rotation.
To address these matters, we estimate an affine transformation -- one that is restricted to uniform scalings -- to best transform the target region to the user's brushed region.
We then extract the translation, scaling, and rotation of each transformation, and use each of these as dependent variables in building a predictive model.

\subsection{Describing the correlated region -- translation}
\label{analysis:centerdist}

\subsubsection{Model Design}
We employed a linear mixed-effect regression model to study what factors – and factor combinations – are predictive of the translational component of the affine transformation. Specifically, the dependent variable corresponds to the Euclidean norm of the translation vector. Similarly to Section \ref{analysis:iou}, we considered all possible interactions among all factors in our initial model, and arrived at the following model:

\begin{align*}
\text{translation} &\sim \mathcal{N}(\mu, \sigma^2) \\
\mu &= \text{vis} + \text{corr} + \text{shape\_l} : \text{background\_l} \\
&\quad + \text{shape\_l} : \text{background\_l} : \text{vis} \\
&\quad + \text{shape\_l} : \text{background\_l} : \text{corr} \\
&\quad + (1 \mid \text{user})
\end{align*}


\subsubsection{Results}
\begin{figure}[!t]
    \centering
    \includegraphics[width=.5\textwidth]{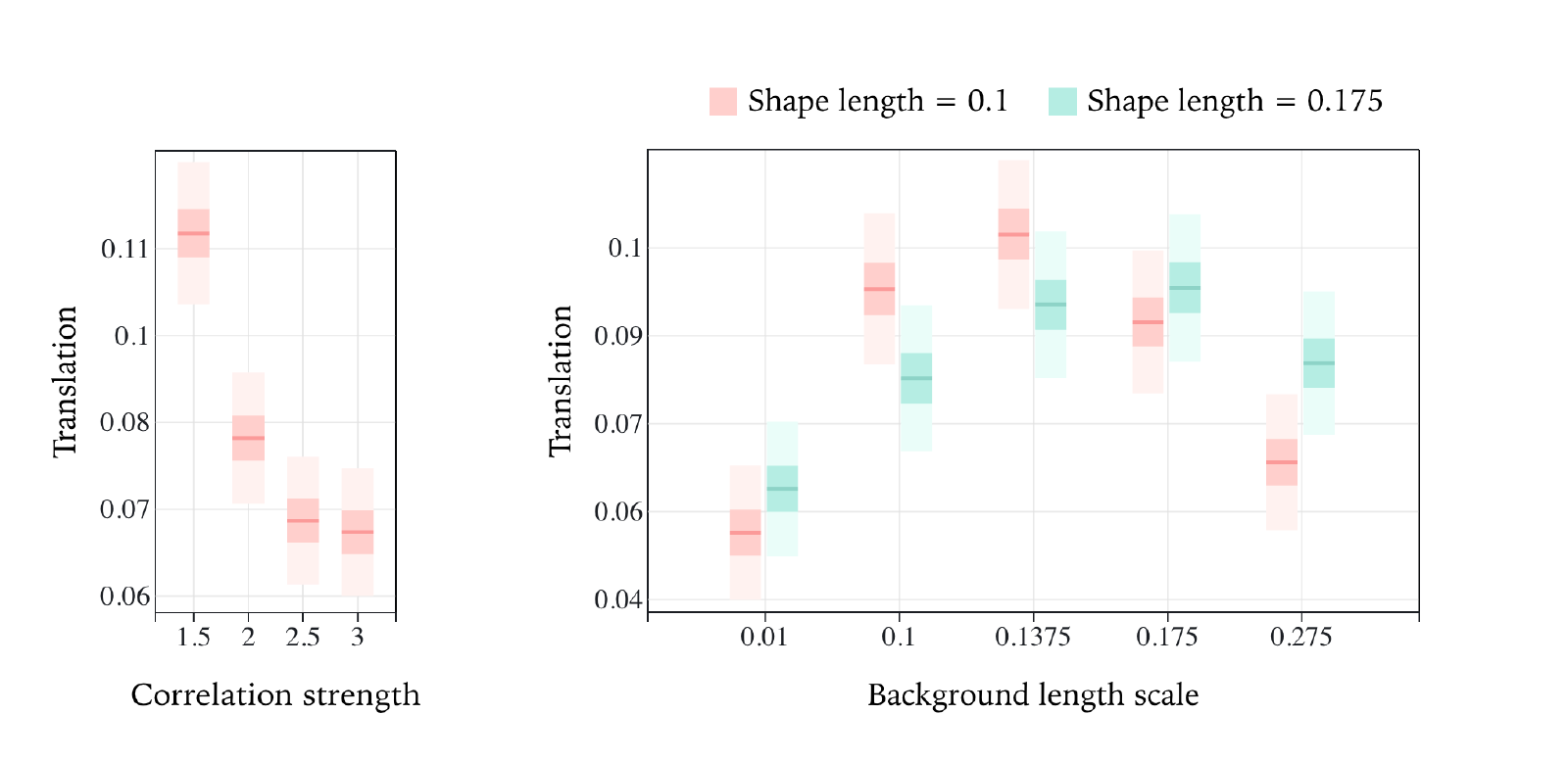}
    \caption{For the effect of translation misalignment in describing the target region, this is most pronounced when the correlation strength is smallest, and the discriminability level is smallest.}
    \label{fig:centerdist:noise:rl}
\end{figure}

\textcolor{midblue}{[corr; shape\_l $\times$ background\_l]} As shown in Fig. \ref{fig:centerdist:noise:rl}, we found that correlation strength ($\chi^2 = 104.1071, Df = 3, p < \num{e-5}$) and the discriminability of strongly correlated region ($\chi^2 = 115.7455, Df = 8, p < \num{e-5}$) 
have significant effects. The higher the correlation strength, the more accurate participants are in estimating the region's location. Similarly, the easier to recognize the region in a single draw, as measured by level of discriminability, the higher the accuracy.

\begin{figure}[!t]
    \centering
    \includegraphics[width=.49\textwidth]{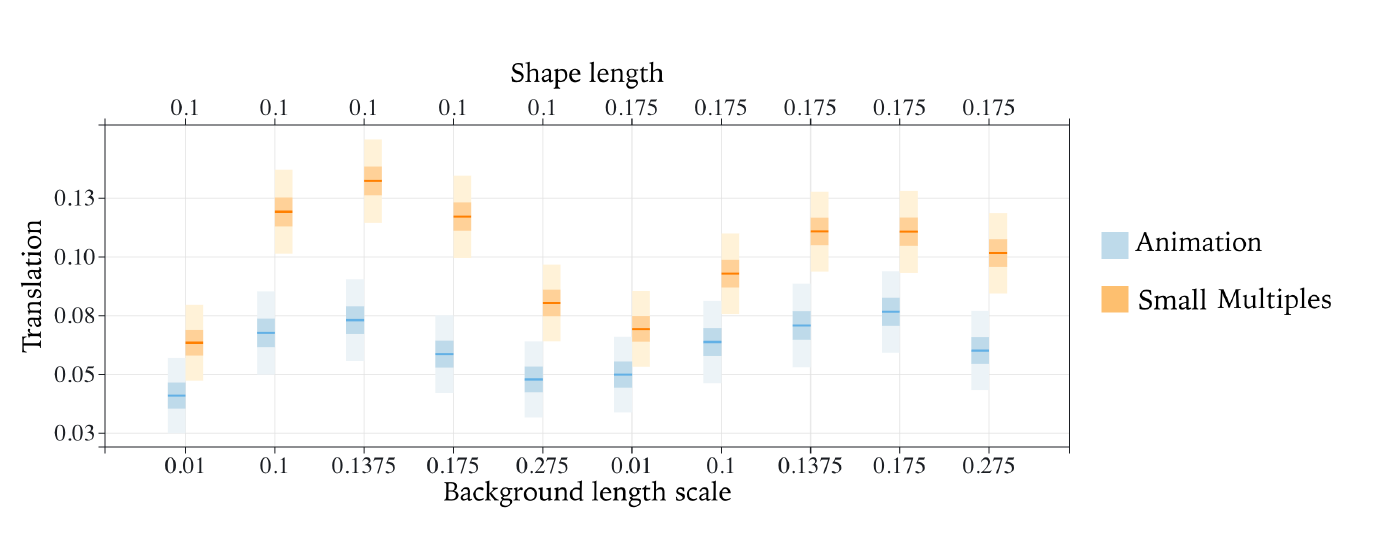}
    \caption{We find that small multiples consistently leads to higher misalignment, as measured by the translational component of an affine transformation between regions. This effect is consistent across the different factors of the data.}
    \label{fig:centerdist:vis_r_l}
\end{figure}

\textcolor{midblue}{[vis ; vis $\times$ shape\_l $\times$ background\_l ; vis $\times$ corr]} We find that small multiples resulted in significantly ($estimate = -0.0396, SE = 0.00868434, Df = 87.1158482, t.ratio = -4.5654638, p =$\num{1.627e-5}) more error in terms of translation compared to animation.
Moreover, this effect is present across both shape length, and background length scale ($\chi^2 = 21.6055, Df = 9, p = 0.01022$), as shown in Fig.~\ref{fig:centerdist:vis_r_l}.
These results imply that translational misalignment plays a role in explaining position-wise misalignment, as measured by IoU.
Therefore, a participant's impression of the target region might be accurate (e.g. the shape), but their specified location of the region might be inaccurate.
We note that error due to translation is not distributed equally across conditions.
The translation increases as the difficulty of the task increases, e.g. lower levels of correlation and discriminability.

\begin{figure}[!t]
    \centering
    \includegraphics[width=.44\textwidth]{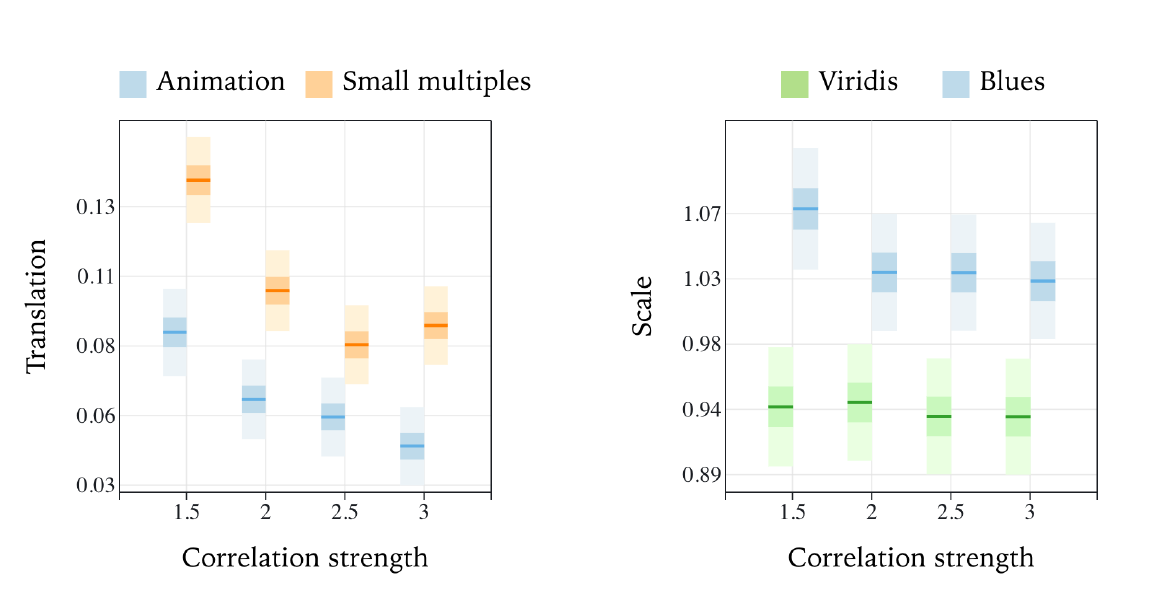}
    \caption{(Left) Small multiples visualization introduces significantly larger error in participants localizing the target region, compared with animation.
    (Right) The Viridis scale will make humans more sensitive to details in the field the compared with the Blues scale, and thus more conservative about strongly correlated region with consistent color.}
    \label{fig:centerdist:vis_noise}
\end{figure}

\subsection{Describing the correlated region -- uniform scaling}
\label{analysis:scale}

\subsubsection{Model Design}
Although the translational component of the affine transformation gives a deeper insight into misalignment, it might not be the only explanation of error.
Thus we next turn to the scaling component of the transformation.
We employ a linear mixed-effect regression model to study what factors – and factor combinations – are predictive of scale, giving:

\begin{align*}
\text{scale} &\sim \mathcal{N}(\mu, \sigma^2) \\
\mu &= \text{color} \times \text{corr} + \text{shape\_1} \times \text{corr} \\
&\quad + \text{shape\_l} : \text{background\_l} \\
&\quad + \text{shape\_l} : \text{background\_l} : \text{corr} \\
&\quad + (1 \mid \text{user})
\end{align*}

\subsubsection{Results}

In our previous analysis, we had found that the inherent spatial alignment issues of small multiples results in significantly more error in how participants perceive the location of the target region.
However, we find that that it does not result in poor judgment on the size of the region.
Nevertheless, we do find scale to have an impact on \textcolor{midblue}{[color]}.
As shown in Fig. \ref{fig:centerdist:vis_noise} (right), we find Viridis not only better helps participants recognize the existence of strongly correlated region, as we discussed in Section \ref{analysis:freq}, but also leads participants to underestimate ($emmean = 0.937,	SE = 0.0197,	Df = 86.083$) the size of the strongly correlated region. Analogously, the Blues color scale leads participants to overestimate ($emmean = 1.043,	SE = 0.0197, Df = 86.66$) the size of the strongly correlated region. As the transformation maps the target region to the brushed region, a value larger than 1 indicates overestimation, while a value less than 1 indicates underestimation.

\textcolor{midblue}{[color $\times$ corr]} We find that the Blues color scale leads to less error in scale compared to Viridis, for all levels of correlation strength except the lowest.
This finding verifies our hypothesis (\textbf{H3}) that multi-hue and single-hue luminance-varying color scales indeed result in varying perception of correlation in 2D scalar field distributions.
The more details and contrast offered in a multi-hue color scale might lead participants to be overly conservative in identifying a constant region of color.

\subsection{Describing the correlated region -- rotation}
\label{analysis:orient}

Lastly, we utilize the estimated rotational component of the affine transformation as a dependent variable, and employed a linear mixed-effect regression model to study if visualization types or color scales will influence human judgment. Through an iterative backward model selection and refinement process, our resulting model indicated no significant effects with respect to visualization type and color scale. This implies that the main effects of transformation are limited to translation (misalignment in small multiples) and a uniform scaling (conservative estimation of shape in Viridis).

%% file: sections/discussion.tex

In this work we investigated the perception of correlation within 2D scalar field distributions.
We contributed a user study to explore the effectiveness of different visualization types, namely animation-based displays and small multiples displays,
as well as different color scales, namely multi-hue and single hue color ramps.
Through our data factors -- correlation strength, discriminability level -- and different types of judgment collected by participants -- recognition of correlation, describing the region of correlation -- our study revealed a number of findings that we summarize and elaborate on below.

We find that even for weakly correlated distributions, participants can recognize the existence of correlation in around $31.8 \pm 2.27\%$ of times through animation displays, and $37.7 \pm 2.45\%$ through small multiple displays.
As for the highest correlation strength (noise = 3), participants can recognize the existence of correlation in around $77.3 \pm 1.84\%$ of times through animation, and $80.6 \pm 1.64\%$ of times through small multiples.
These findings suggest that simple and direct displays of 2D scalar field distributions have the ability to support visual analysis of such data.
This stands in contrast with other, more complex visualizations based on derived spatial features, which might not be strictly necessary for analysis.

In addition, we find a small multiples display can better help participants recognize correlation in comparison to animation, provided the strongly correlated region is sufficiently large.
For such large target regions, we hypothesize that the time constraints of animation can lead to uncertainty in participants on whether some observed views consist of a coherent, recognizable region.
In contrast, small multiples displays give users the time to analyze individual draws, and in turn relate and ensemble the information.
In isolation, one view might not appear to contain the target region, but when integrating the remaining views, one might be able to find commonality in visual patterns and arrive at a different conclusion.

We further investigated how different visualization designs influence the accuracy of their perception of correlated regions.
For example, although small multiples can better help humans recognize the existence of correlation, it suffers from an inherent spatial alignment issue, and will introduce larger error to human perception and summary of the correlated region.
Such a position-wise measure of misalignment, however, does not reveal \emph{why} such error exists.
Through explicitly finding a transformation to align a target region with the user's brushed region, we may decompose error along translation, uniform scaling, and rotation.
Under this perspective, we find that small multiples largely introduces errors in the location of the correlated region in comparison with animation, instead of the region's orientation or size.
This implies that the higher recognition in small multiples is not merely an illusion, but rather, participants accurately perceive the target region, and only incur error in precisely localizing the region.
Moreover, this decomposition further highlights effects in color, revealing that the Viridis color scale leads participants to underestimate size, in comparison to the Blues color scale.
This demonstrates a trade-off in correlation recognition: although the additional details in Viridis likely leads to higher recognition of the target region, it comes at a cost of making overly conservative estimates in describing the region.

\subsection{Limitations and future work}

We acknowledge several limitations with our work.
Our study is a first step towards evaluating perceptual judgments of spatial correlation in 2D scalar field distributions. As such, we limit the task to the simplest kind of correlation pattern, where the correlated region is distinguished by having a constant value within a single draw. In relaxing this condition towards more complex correlation patterns (c.f. Fig.~\ref{fig:height_field}-top), we believe our experimental design can serve as a basis for future studies.
An additional factor in limiting our experimental design is the consideration of just two color maps: Viridis and Blues.
Although we believe these are representative of, respectively, multi-hue and single-hue luminance varying color scales, numerous other choices for sequential color scales exist, e.g. a fully desaturated color scale, or using a rainbow-style color scale~\cite{reda2022rainbow} to better emphasize details.
We note that our experimental design does not prohibit the study of such color scales, but rather, we precluded other color scales to reduce the potential size of the study.

Furthermore, we found it necessary to manually set a number of parameters between visualization conditions.
As discussed in Sec.~\ref{sec:interface}, parameters were set to maximize the respective strengths of both visualizations, given a limited time budget to complete the task.
A consequence, however, is that the conditions are assigned different data.
If we relaxed the time constraint, it would be possible to address this limitation, as we could allow for users to casually browse a large gallery of views in the small multiples design, potentially with the aid of scrubbing interaction tools.
Ultimately we opted against this experimental design as the increased flexibility could introduce high inter-participant variability.
Furthermore, we decided on a simple way to display animated frames, whereas smooth inter-frame transitions could have potentially eased the task difficulty.


Beyond addressing the aforementioned limitations, we see a number of directions for future work.
We intend to expand our data generation framework towards more realistic settings.
At the moment, we chose simple shapes for the target region -- ellipses -- to more easily control for the factor of shape.
However, for ecological validity, shapes that are more reflective of real-world correlated regions should be considered, e.g. correlation structures found in climate-based simulation ensembles.
Our method of controlling for shape, namely specification of the covariance matrix of a Gaussian, does not inhibit the specification of more complex shapes, but rather, an adjustment to our foreground model.
Moreover, although our study was limited to the use of color maps, we plan to investigate other forms of scalar field visualizations.
Specifically, contour-based visualizations can give another perspective on correlation, whether shown directly, or aggregated in a single view, e.g. as in contour box plots~\cite{whitaker2013contour}.

%% file: vis.bbl
\begin{thebibliography}{10}

\bibitem{potter2010visualizing}
Visualizing summary statistics and uncertainty.
\newblock In {\em Computer Graphics Forum}, vol.~29, pp. 823--832. Wiley Online Library, 2010.

\bibitem{archambault2010animation}
D.~Archambault, H.~Purchase, and B.~Pinaud.
\newblock Animation, small multiples, and the effect of mental map preservation in dynamic graphs.
\newblock {\em IEEE transactions on visualization and computer graphics}, 17(4):539--552, 2010.

\bibitem{banerjee2008gaussian}
S.~Banerjee, A.~E. Gelfand, A.~O. Finley, and H.~Sang.
\newblock Gaussian predictive process models for large spatial data sets.
\newblock {\em Journal of the Royal Statistical Society Series B: Statistical Methodology}, 70(4):825--848, 2008.

\bibitem{beecham2016map}
R.~Beecham, J.~Dykes, W.~Meulemans, A.~Slingsby, C.~Turkay, and J.~Wood.
\newblock Map lineups: Effects of spatial structure on graphical inference.
\newblock {\em IEEE transactions on visualization and computer graphics}, 23(1):391--400, 2016.

\bibitem{boukhelifa2023visualization}
N.~Boukhelifa, C.~R. Johnson, and K.~Potter.
\newblock Visualization and decision making design under uncertainty.
\newblock {\em IEEE Computer Graphics and Applications}, 43(5):23--25, 2023.

\bibitem{boyandin2012qualitative}
I.~Boyandin, E.~Bertini, and D.~Lalanne.
\newblock A qualitative study on the exploration of temporal changes in flow maps with animation and small-multiples.
\newblock In {\em Computer Graphics Forum}, vol.~31, pp. 1005--1014. Wiley Online Library, 2012.

\bibitem{brehmer2019comparative}
M.~Brehmer, B.~Lee, P.~Isenberg, and E.~K. Choe.
\newblock A comparative evaluation of animation and small multiples for trend visualization on mobile phones.
\newblock {\em IEEE Transactions on Visualization and Computer Graphics}, 26(1):364--374, 2019.

\bibitem{bujack2017good}
R.~Bujack, T.~L. Turton, F.~Samsel, C.~Ware, D.~H. Rogers, and J.~Ahrens.
\newblock The good, the bad, and the ugly: A theoretical framework for the assessment of continuous colormaps.
\newblock {\em IEEE transactions on visualization and computer graphics}, 24(1):923--933, 2017.

\bibitem{bykov2021explaining}
K.~Bykov, M.~M.-C. H{\"o}hne, A.~Creosteanu, K.-R. M{\"u}ller, F.~Klauschen, S.~Nakajima, and M.~Kloft.
\newblock Explaining bayesian neural networks.
\newblock {\em arXiv preprint arXiv:2108.10346}, 2021.

\bibitem{correll2014error}
M.~Correll and M.~Gleicher.
\newblock Error bars considered harmful: Exploring alternate encodings for mean and error.
\newblock {\em IEEE transactions on visualization and computer graphics}, 20(12):2142--2151, 2014.

\bibitem{farrugia2011effective}
M.~Farrugia and A.~Quigley.
\newblock Effective temporal graph layout: A comparative study of animation versus static display methods.
\newblock {\em Information Visualization}, 10(1):47--64, 2011.

\bibitem{ferstl2015streamline}
F.~Ferstl, K.~B{\"u}rger, and R.~Westermann.
\newblock Streamline variability plots for characterizing the uncertainty in vector field ensembles.
\newblock {\em IEEE Transactions on Visualization and Computer Graphics}, 22(1):767--776, 2015.

\bibitem{ferstl2016visual}
F.~Ferstl, M.~Kanzler, M.~Rautenhaus, and R.~Westermann.
\newblock Visual analysis of spatial variability and global correlations in ensembles of iso-contours.
\newblock In {\em Computer Graphics Forum}, vol.~35, pp. 221--230. Wiley Online Library, 2016.

\bibitem{harrison2014ranking}
L.~Harrison, F.~Yang, S.~Franconeri, and R.~Chang.
\newblock Ranking visualizations of correlation using weber's law.
\newblock {\em IEEE transactions on visualization and computer graphics}, 20(12):1943--1952, 2014.

\bibitem{holalkere2025stochastic}
S.~Holalkere, D.~Bindel, S.~Sell{\'a}n, and A.~Terenin.
\newblock Stochastic poisson surface reconstruction with one solve using geometric gaussian processes.
\newblock In {\em Forty-second International Conference on Machine Learning}, 2025.

\bibitem{hollt2014ovis}
T.~H{\"o}llt, A.~Magdy, P.~Zhan, G.~Chen, G.~Gopalakrishnan, I.~Hoteit, C.~D. Hansen, and M.~Hadwiger.
\newblock Ovis: A framework for visual analysisof ocean forecast ensembles.
\newblock {\em IEEE Transactions on Visualization and Computer Graphics}, 20(8):1114--1126, 2014.

\bibitem{hosseinpour2024examining}
H.~Hosseinpour, L.~E. Matzen, K.~M. Divis, S.~C. Castro, and L.~Padilla.
\newblock Examining limits of small multiples: Frame quantity impacts judgments with line graphs.
\newblock {\em IEEE Transactions on Visualization and Computer Graphics}, 2024.

\bibitem{hullman2018pursuit}
J.~Hullman, X.~Qiao, M.~Correll, A.~Kale, and M.~Kay.
\newblock In pursuit of error: A survey of uncertainty visualization evaluation.
\newblock {\em IEEE transactions on visualization and computer graphics}, 25(1):903--913, 2018.

\bibitem{hullman2015hypothetical}
J.~Hullman, P.~Resnick, and E.~Adar.
\newblock Hypothetical outcome plots outperform error bars and violin plots for inferences about reliability of variable ordering.
\newblock {\em PloS one}, 10(11):e0142444, 2015.

\bibitem{ibrekk1987graphical}
H.~Ibrekk and M.~G. Morgan.
\newblock Graphical communication of uncertain quantities to nontechnical people.
\newblock {\em Risk analysis}, 7(4):519--529, 1987.

\bibitem{kale2018hypothetical}
A.~Kale, F.~Nguyen, M.~Kay, and J.~Hullman.
\newblock Hypothetical outcome plots help untrained observers judge trends in ambiguous data.
\newblock {\em IEEE transactions on visualization and computer graphics}, 25(1):892--902, 2018.

\bibitem{kay2015beyond}
M.~Kay and J.~Heer.
\newblock Beyond weber's law: A second look at ranking visualizations of correlation.
\newblock {\em IEEE transactions on visualization and computer graphics}, 22(1):469--478, 2015.

\bibitem{kay2016ish}
M.~Kay, T.~Kola, J.~R. Hullman, and S.~A. Munson.
\newblock When (ish) is my bus? user-centered visualizations of uncertainty in everyday, mobile predictive systems.
\newblock In {\em Proceedings of the 2016 chi conference on human factors in computing systems}, pp. 5092--5103, 2016.

\bibitem{liu2018somewhere}
Y.~Liu and J.~Heer.
\newblock Somewhere over the rainbow: An empirical assessment of quantitative colormaps.
\newblock In {\em Proceedings of the 2018 CHI conference on human factors in computing systems}, pp. 1--12, 2018.

\bibitem{mackay1998introduction}
D.~J. MacKay et~al.
\newblock Introduction to gaussian processes.
\newblock {\em NATO ASI series F computer and systems sciences}, 168:133--166, 1998.

\bibitem{maddox2021bayesian}
W.~J. Maddox, M.~Balandat, A.~G. Wilson, and E.~Bakshy.
\newblock Bayesian optimization with high-dimensional outputs.
\newblock {\em Advances in neural information processing systems}, 34:19274--19287, 2021.

\bibitem{mateevitsi2024science}
V.~A. Mateevitsi, M.~E. Papka, and K.~Reda.
\newblock Science in a blink: Supporting ensemble perception in scalar fields.
\newblock {\em arXiv preprint arXiv:2406.14452}, 2024.

\bibitem{munzner2014visualization}
T.~Munzner.
\newblock {\em Visualization analysis and design}.
\newblock CRC press, 2014.

\bibitem{padilla2020uncertainty}
L.~Padilla, M.~Kay, and J.~Hullman.
\newblock Uncertainty visualization.
\newblock 2020.

\bibitem{palmer1999vision}
S.~E. Palmer.
\newblock {\em Vision science: Photons to phenomenology}.
\newblock MIT press, 1999.

\bibitem{pfaffelmoser2012visualization}
T.~Pfaffelmoser and R.~Westermann.
\newblock Visualization of global correlation structures in uncertain 2d scalar fields.
\newblock In {\em Computer Graphics Forum}, vol.~31, pp. 1025--1034. Wiley Online Library, 2012.

\bibitem{pont2021wasserstein}
M.~Pont, J.~Vidal, J.~Delon, and J.~Tierny.
\newblock Wasserstein distances, geodesics and barycenters of merge trees.
\newblock {\em IEEE Transactions on Visualization and Computer Graphics}, 28(1):291--301, 2021.

\bibitem{reckinger2015study}
S.~M. Reckinger, M.~R. Petersen, and S.~J. Reckinger.
\newblock A study of overflow simulations using mpas-ocean: Vertical grids, resolution, and viscosity.
\newblock {\em Ocean Modelling}, 96:291--313, 2015.

\bibitem{reda2022rainbow}
K.~Reda.
\newblock Rainbow colormaps: What are they good and bad for?
\newblock {\em IEEE Transactions on Visualization and Computer Graphics}, 29(12):5496--5510, 2022.

\bibitem{reda2020rainbows}
K.~Reda and D.~A. Szafir.
\newblock Rainbows revisited: Modeling effective colormap design for graphical inference.
\newblock {\em IEEE transactions on visualization and computer graphics}, 27(2):1032--1042, 2020.

\bibitem{robertson2008effectiveness}
G.~Robertson, R.~Fernandez, D.~Fisher, B.~Lee, and J.~Stasko.
\newblock Effectiveness of animation in trend visualization.
\newblock {\em IEEE transactions on visualization and computer graphics}, 14(6):1325--1332, 2008.

\bibitem{sanyal2010noodles}
J.~Sanyal, S.~Zhang, J.~Dyer, A.~Mercer, P.~Amburn, and R.~Moorhead.
\newblock Noodles: A tool for visualization of numerical weather model ensemble uncertainty.
\newblock {\em IEEE transactions on visualization and computer graphics}, 16(6):1421--1430, 2010.

\bibitem{sarma2022evaluating}
A.~Sarma, S.~Guo, J.~Hoffswell, R.~Rossi, F.~Du, E.~Koh, and M.~Kay.
\newblock Evaluating the use of uncertainty visualisations for imputations of data missing at random in scatterplots.
\newblock {\em IEEE Transactions on Visualization and Computer Graphics}, 29(1):602--612, 2022.

\bibitem{sellan2022stochastic}
S.~Sell{\'a}n and A.~Jacobson.
\newblock Stochastic poisson surface reconstruction.
\newblock {\em ACM Transactions on Graphics (TOG)}, 41(6):1--12, 2022.

\bibitem{srabanti2022comparative}
S.~Srabanti, C.~Veiga, E.~Silva, M.~Lage, N.~Ferreira, and F.~Miranda.
\newblock A comparative study of methods for the visualization of probability distributions of geographical data.
\newblock {\em Multimodal Technologies and Interaction}, 6(7):53, 2022.

\bibitem{szafir2017modeling}
D.~A. Szafir.
\newblock Modeling color difference for visualization design.
\newblock {\em IEEE transactions on visualization and computer graphics}, 24(1):392--401, 2017.

\bibitem{todd2004capacity}
J.~J. Todd and R.~Marois.
\newblock Capacity limit of visual short-term memory in human posterior parietal cortex.
\newblock {\em Nature}, 428(6984):751--754, 2004.

\bibitem{tufte1983visual}
E.~R. Tufte and P.~R. Graves-Morris.
\newblock {\em The visual display of quantitative information}, vol.~2.
\newblock Graphics press Cheshire, CT, 1983.

\bibitem{um2021spot}
K.~Um, X.~Hu, B.~Wang, and N.~Thuerey.
\newblock Spot the difference: Accuracy of numerical simulations via the human visual system.
\newblock {\em ACM Transactions on Applied Perception (TAP)}, 18(2):1--15, 2021.

\bibitem{ware2019information}
C.~Ware.
\newblock {\em Information visualization: perception for design}.
\newblock Morgan Kaufmann, 2019.

\bibitem{whitaker2013contour}
R.~T. Whitaker, M.~Mirzargar, and R.~M. Kirby.
\newblock Contour boxplots: A method for characterizing uncertainty in feature sets from simulation ensembles.
\newblock {\em IEEE Transactions on Visualization and Computer Graphics}, 19(12):2713--2722, 2013.

\bibitem{williams2006gaussian}
C.~K. Williams and C.~E. Rasmussen.
\newblock {\em Gaussian processes for machine learning}, vol.~2.
\newblock MIT press Cambridge, MA, 2006.

\bibitem{wilson2020efficiently}
J.~Wilson, V.~Borovitskiy, A.~Terenin, P.~Mostowsky, and M.~Deisenroth.
\newblock Efficiently sampling functions from gaussian process posteriors.
\newblock In {\em International Conference on Machine Learning}, pp. 10292--10302. PMLR, 2020.

\bibitem{zhang2021visualizing}
D.~Zhang, E.~Adar, and J.~Hullman.
\newblock Visualizing uncertainty in probabilistic graphs with network hypothetical outcome plots (nethops).
\newblock {\em IEEE Transactions on Visualization and Computer Graphics}, 28(1):443--453, 2021.

\end{thebibliography}
